\documentclass[12pt]{iopart}
\usepackage{iopams}
\usepackage{amsopn}
\usepackage{graphicx}
\usepackage{floatflt}
\usepackage{bm}
\usepackage{cite}

\DeclareMathOperator{\res}{res} 
\DeclareMathOperator{\ad}{ad} 
\DeclareMathOperator{\Ad}{Ad} \DeclareMathOperator{\Mat}{Mat}
\DeclareMathOperator{\diag}{diag} \DeclareMathOperator{\Ker}{Ker}

\begin{document}

\title[A generalised Landau-Lifshitz equation for isotropic
SU(3) magnet]{A generalised Landau-Lifshitz equation for isotropic
SU(3) magnet}

\author{J Bernatska$^{1,2}$ and P Holod$^{1,2}$ }

\address{${}^1$ National University of `Kiev-Mohyla Academy', 2
Skovorody Str., 04070 Kiev, UA}

\address{${}^2$ Bogolyubov Institute for Theoretical Physics, 14b
Metrologichna Str., 03680 Kiev, UA}
\ead{\mailto{BernatskaJM@ukma.kiev.ua}, \mailto{Holod@ukma.kiev.ua}}

\begin{abstract}
In the paper we obtain equations for large-scale fluctuations of the
mean field (the field of magnetization and quadrupole moments) in a
magnetic system realized by a square (cubic) lattice of atoms with
spin $s\,{\geqslant }\,1$ at each site. We use the generalized
Heisenberg Hamiltonian with biquadratic exchange as a quantum model.
A quantum thermodynamical averaging gives classical effective
models, which are interpreted as Hamiltonian systems on coadjoint
orbits of Lie group SU(3).
\end{abstract}


\submitto{\JPA}
\maketitle

\section{Introduction}
Being a multiparticle quantum system, a magnet can be considered on
different levels of hierarchy: a quantum (microscopic) level and a
classical (macroscopic) one. The~quantum level is described by means
of quantum electrodynamics, or by simpler models like the Hubbard
model or the Heisenberg one. The most common model for the classical
level is the mean field model. Dynamics of a mean field is described
by the equations of Landau-Lifshitz type.

Each model is suitable to describe certain phenomena. For example,
the problems of formation of large-scale structures (domain walls,
topological solitons, nonlinear magnetization waves and so on) are
naturally investigated from a classical point of view. More tenuous
problems, like renormalization of the order parameter according to a
temperature or an effective interaction constant, require a quantum
point of view~\cite{Tsvelyk}.

Here we start from the quantum level described by the Heisenberg
model. In~addition to the usual Heisenberg bilinear interaction
$-J(\hat{\bi{S}}_n,\hat{\bi{S}}_{m})$, we consider the biquadratic
one $-K(\hat{\bi{S}}_n,\hat{\bi{S}}_{m})^2$. By many theoretical and
experimental researches it was shown that the biquadratic
interactions have significant effects on magnetic properties. For
example, a new ordered state (a nematic state, with zero
magnetization) occurs as a separate phase transition~\cite{Blume}.
Note, that the biquadratic interaction can be taken into account
only if a magnetic system has the spin $s\,{\geqslant}\,1$.

In this paper we propose a classical generalization of the isotropic
Landau-Lifshitz equation corresponding to the Heisenberg model with
biquadratic exchange interaction. A transition from the quantum
level to the classical one is performed by the mean field
approximation. The classical model can be interpreted as a
Hamiltonian system on a coadjoint orbit of the unitary group
$\mathrm{SU}(3)$. Therefore, we acquire an additional mathematical
apparatus, which gives a significant advantage.

The mean field approximation gives a qualitative analysis of ordered
states \cite{Matveev, Nauciel}, but has no answer about their
stability. Moreover, in this approximation the temperature
dependencies of order parameters  considerably differ from the
observed dependencies. That proves a necessity to take into account
fluctuations of the mean field. The~proposed effective classical
models describe large-scale (or slow) fluctuations of mean field.
One~can come to slow fluctuations by an averaging over high
frequencies~\cite{Tsvelyk}. However, remaining in the context of
theory of magnetism, we choose the models associated with the
equations of Landau-Lifshitz type.

The paper is organized as follows. Section 2 is devoted to the
quantum model based on the spin Hamiltonian with biquadratic
exchange interactions. We consider the SU(3)-invariant case. In
section 3 we construct two effective models that describe
large-scale fluctuations of mean field (the field of magnetization
and quadrupole moments). We~obtain one of them by an averaging of
the quantum Hamiltonian over coherent states. The other effective
model is a result of an averaging over mixed states. These classical
models appear to be Hamiltonian systems on coadjoint orbits of the
group SU(3), that follows from SU(3)-invariance of the original
quantum model. Each coadjoint orbit is determined by constrains,
which are observed quantities becoming rigid after averaging. In
section 4 we summarize results and give some ideas how to extend the
proposed scheme to magnetic systems with higher spins.

\section{Quantum model of magnetic system}
\subsection{Description of the model}
The magnetic system in question is realized by a homogeneous lattice
of atoms with the spin $s\,{\geqslant}\,1$ at each site. The lattice
can be one-, two-, or three-dimensional, and has the distance $l$
between the nearest-neighbor sites. We assign three \emph{spin
operators} $(\hat{S}^{1}_n,\,\hat{S}^{2}_n, \,\hat{S}^{3}_n)$ to
each site $n$; they obey the standard commutation relations:
\begin{equation*}
[\hat{S}_{n}^{\alpha},\hat{S}_{m}^{\beta}] =
\rmi\varepsilon^{\alpha\beta\gamma} \hat{S}_{n}^{\gamma}
\delta_{nm},
\end{equation*}
where $\alpha$, $\beta$, $\gamma$ run over the set $\{1,\,2,\,3\}$,
and $\delta_{nm}$ denotes the Kronecker symbol.

We use the localized spin model for the magnetic system. In many
cases this model adequately describes a magnetic system by the
Heisenberg Hamiltonian, which includes only the bilinear exchange
interaction. Nevertheless, there are a lot of magnets that require
taking into account higher powers of exchange interaction. Our model
is applicable to magnets with the spin $s\,{\geqslant}\, 1$.

In the present paper we consider the Hamiltonian with biquadratic
exchange and call it \emph{bilinear-biquadratic}:
\begin{equation}\label{BBH}
\hat{\mathcal{H}} = -\sum_{n,\delta}
\{J(\hat{\bi{S}}_n,\hat{\bi{S}}_{n+\delta})+K
(\hat{\bi{S}}_n,\hat{\bi{S}}_{n+\delta})^2\},
\end{equation}
where $\hat{\bi{S}}_n = (\hat{S}_n^1,\,\hat{S}_n^2,\,\hat{S}_n^3)$
is a vector of spin operators at site $n$, and $\delta$ runs over
the nearest-neighbor sites.  This Hamiltonian was discussed, for
example, in \cite{Blume,Matveev,Nauciel,Buchta}. The constants $J$
and $K$ serve as exchange integrals. We suppose that $J$ and $K$ are
positive. It means that we consider a ferromagnetic interaction in
preference.

The operators $\{\hat{S}_n^{\alpha}\}$ (here $n$ is fixed) are
defined over the $(2s\,{+}\,1)$-dimensional space of irreducible
representation of the group SU(2). They generate an associative
matrix algebra over this space. The complete matrix algebra can be
represented as a direct sum of irreducible sets of tensor operators
with respect to the action $\ad_{\hat{S}^{\alpha}}$. In the case of
$s\,{=}\,1$, we have: $\Mat_{3\times 3} \,{\simeq}\, [9] \,{=}\,
[1]\,{+}\,[3]\,{+}\,[5]$. Evidently, the operators
$\{\hat{S}^{\alpha}_n\}$  form a basis in the 3-dimensional
irreducible set. One can construct a basis in the 5-dimensional
irreducible set from the tensor operators of weight 2. These are the
\emph{quadrupole operators}
$\{\hat{Q}_n^{12},\,\hat{Q}_n^{13},\,\hat{Q}_n^{23}$,
$\hat{Q}_n^{[2,2]}$, $\hat{Q}_n^{[2,0]}\}$ defined by the formulas:
\begin{eqnarray*}
\hat{Q}_n^{\alpha\beta} = \hat{S}_n^{\alpha} \hat{S}_n^{\beta} +
\hat{S}_n^{\beta} \hat{S}_n^{\alpha},\ \alpha\neq\beta,\\
\hat{Q}_n^{[2,2]}=(\hat{S}_n^1)^2-(\hat{S}_n^2)^2, \quad
\hat{Q}_n^{[2,0]}=\sqrt{3}\Bigl((\hat{S}_n^3)^2-\textstyle\frac{2}{3}\Bigr).
\end{eqnarray*}
The spin and quadrupole operators are normalized by the following
relation:
\begin{equation*}
\Tr (\hat{P})^2 = \textstyle\frac{1}{3} s(s+1) (2s+1).
\end{equation*}
As $s\,{=}\,1$ we have  $\Tr (\hat{P})^2 \,{=}\, 2$ . The chosen
normalization is matched to the relation $(\hat{S}_n^1)^2 +
(\hat{S}_n^2)^2 + (\hat{S}_n^3)^2 \,{=}\, s(s+1)$.

Now, fix the canonical basis $\{|{+}1\rangle,\, |{-}1\rangle,\,
|0\rangle\}$ in the space of representation. Then one obtains the
following matrix representation for the spin and quadrupole
operators:
\begin{eqnarray*}
 \hat{S}_n^1=\frac{1}{\sqrt{2}}\left(\begin{array}{ccc}
 0 & 0 & 1 \\ 0 & 0 & 1 \\ 1 & 1 & 0  \end{array}\right),\
 \hat{S}_n^2=\frac{1}{\sqrt{2}}\left(\begin{array}{ccc}
 0 & 0 & -\rmi \\ 0 & 0 & \rmi \\ \rmi & -\rmi & 0
 \end{array}\right),\\
 \hat{S}_n^3=\left(\begin{array}{ccc}
 1 & 0 & 0 \\ 0 & -1 & 0 \\ 0 & 0 & 0
 \end{array}\right), \
 \hat{Q}_n^{[2,0]}=\frac{1}{\sqrt{3}}\left(\begin{array}{ccc}
 1 & 0 & 0 \\ 0 & 1 & 0 \\ 0 & 0 & -2
 \end{array}\right),\\
 \hat{Q}_n^{12}=\left(\begin{array}{ccc}
 0 & -\rmi & 0 \\ \rmi & 0 & 0 \\ 0 & 0 & 0
 \end{array}\right),\
 \hat{Q}_n^{13}=\frac{1}{\sqrt{2}}\left(\begin{array}{ccc}
 0 & 0 & 1 \\ 0 & 0 & -1 \\ 1 & -1 & 0
 \end{array}\right),\\
 \hat{Q}_n^{23}=\frac{1}{\sqrt{2}}\left(\begin{array}{ccc}
 0 & 0 & -\rmi \\ 0 & 0 & -\rmi \\ \rmi & \rmi & 0
 \end{array}\right),\
 \hat{Q}_n^{[2,2]}=\left(\begin{array}{ccc}
 0 & 1 & 0 \\ 1 & 0 & 0 \\ 0 & 0 & 0
 \end{array}\right).
\end{eqnarray*}
We denote all spin and quadrupole operators: $\{\hat{S}_n^{1}$,
$\hat{S}_n^{2}$, $\hat{S}_n^{3}$, $\hat{Q}_n^{12}$,
$\hat{Q}_n^{13}$, $\hat{Q}_n^{23}$, $\hat{Q}_n^{[2,2]}$,
$\hat{Q}_n^{[2,0]}\}$ by $\{\hat{P}^{a}_n\}_{a=1}^8$. The operators
$\{\hat{P}^{a}_n\}$ obey the following commutation relations:
\begin{equation*}
 [\hat{P}^{a}_n,\,\hat{P}^{b}_m] = \rmi C_{abc} \hat{P}^{c}_{n}
 \delta_{nm},
\end{equation*}
where $C_{abc}$ are structure constants; nonzero components are
\begin{eqnarray*}
&C_{123}=C_{145}=C_{167}=C_{264}=C_{257}=C_{356}=1,\\
&C_{168}= C_{528}=\sqrt{3}, \quad C_{437}=2.
\end{eqnarray*}

The Hamiltonian \eref{BBH} becomes bilinear in the terms of
$\{\hat{P}^{a}_n\}$:
\begin{equation}\label{BBH_SU2}
    \hat{\mathcal{H}}=-(J-{\textstyle\frac{1}{2}}K)\sum_{n,\delta}\sum_{\alpha}
    \hat{S}^{\alpha}_n \hat{S}^{\alpha}_{n+\delta}-{\textstyle\frac{1}{2}}K\sum_{n,\delta}
    \sum_{a}\hat{Q}^{a}_{n} \hat{Q}^{a}_{n+\delta}-{\textstyle\frac{4}{3}}KN,
\end{equation}
where $N$ denotes the total number of sites. Obviously, the
Hamiltonian is SU(2)-invariant, and one can transform the operators
$\{\hat{S}_n^{\alpha}\}$ and $\{\hat{Q}_n^a\}$ by the formulas of
adjoint representation
\begin{eqnarray*}
\hat{U}\hat{S}_n^{\alpha} \hat{U}^{-1} = \sum_{\beta}
\hat{D}^{\alpha\beta} (\hat{U})\hat{S}_n^{\beta},\quad
\hat{D}^{\alpha\beta} \in
  \mathrm{SO}(3),\\
\hat{U}\hat{Q}_n^{a} \hat{U}^{-1} = \sum_{b} \hat{D}^{ab}
(\hat{U})\hat{Q}_n^{b}, \quad  \hat{D}^{ab} \in
  \mathrm{SO}(5),
\end{eqnarray*}
where $\hat{D}^{\alpha\beta} (\hat{U})$ and $\hat{D}^{ab} (\hat{U})$
are matrices of the real irreducible 3- and 5-dimensional
representations of the group SU(2) respectively, and $\hat{U}\,{=}\,
\exp\{\sum_{\alpha} \varphi_{\alpha} \hat{S}_n^{\alpha}\}$, where
$\{\varphi_{\alpha}\}$ are group parameters. As $K\,{=}\,J$ the
SU(2)-symmetry is extended to the SU(3)-one, and the Hamiltonian
\eref{BBH_SU2} gets the form
\begin{equation}\label{BBH_SU3}
    \hat{\mathcal{H}}=-{\textstyle\frac{1}{2}}J \sum_{n,\delta}
    \sum_{a}\hat{P}^{a}_{n} \hat{P}^{a}_{n+\delta}-{\textstyle\frac{4}{3}}JN.
\end{equation}

\subsection{Mean field approach and ordered states}\label{ss:MF}
Instead of interactions between the spin and quadrupole operators
$\{\hat{P}^a_n\}$ according to the Hamiltonian \eref{BBH_SU2}, we
consider effective interactions of the operators $\{\hat{P}^a_n\}$
with a classical mean field. We suppose that components of the mean
field at site $n$ are proportional to averages (quasiaverages) of
the quantum operators $\{\hat{P}^a_n\}$.

In the mean field approximation the Hamiltonian \eref{BBH_SU2} has
the form
\begin{equation}\label{HamiltMidField}
    \hat{\mathcal{H}}_{\rm MF}=-(J-{\textstyle\frac{1}{2}}K)z\sum_{n}\sum_{\alpha} \hat{S}_{n}^{\alpha}\langle
    \hat{S}_{n}^{\alpha}\rangle -{\textstyle\frac{1}{2}}Kz\sum_{n}\sum_{a} \hat{Q}_{n}^{a}\langle
    \hat{Q}_{n}^{a}\rangle-{\textstyle\frac{4}{3}}KNz,
\end{equation}
where $z$ is a number of the nearest-neighbor sites. We have to give
a warning about averages of $\{\hat{P}_{n}^{\alpha}\}$. If one
calculates the averages by means of the density matrix
$\hat{\rho}(T) \,{=}\, \exp \{-\frac{\mathcal{H}}{kT}\}$, one
obtains zeros. This follows from the SU(2)-symmetry of the
Hamiltonian \eref{BBH_SU2}. Nonzero values of the averages appear if
the symmetry is broken. Symmetry breaking can be stimulated by an
external magnetic field that vanishes after specifying an order in
the magnetic system. Such averages are called \emph{quasiaverages}
\cite{Bogolyubov}.

Suppose that the magnetic system in question has nonzero
quasiaverages $\{\langle \hat{P}_{n}^{\alpha}\rangle\}$. They
 form a classical 8-component vector field
$\{\mu_a(\bi{x}_n)\}_{a=1}^8$, which we call a \emph{mean field}.
Suppose that the mean field is constant over the whole magnetic
system. This happens in the case of thermodynamic equilibrium and an
infinite lattice. Then under an action of the group SU(2) the
Hamiltonian \eref{HamiltMidField} can be reduced to a diagonal form,
namely:
\begin{eqnarray*}
   \hat{\mathcal{H}}_{\rm MF}&=-(J-{\textstyle\frac{1}{2}}K)z\sum_{n}\hat{S}_{n}^{3}\langle
    \hat{S}_{n}^{3}\rangle -{\textstyle\frac{1}{2}}Kz\sum_{n}\hat{Q}_{n}^{[2,0]}\langle
    \hat{Q}_{n}^{[2,0]}\rangle-{\textstyle\frac{4}{3}}KNz =\\ &= -z\sum_{n} \left\{(J-{\textstyle\frac{1}{2}}K)
    \hat{S}_n^{3}\,\mu_3+{\textstyle\frac{1}{2}}K  \hat{Q}_n^{[2,0]}\mu_8+
    {\textstyle\frac{4}{3}}K\right\},
\end{eqnarray*}
where the components $\mu_3\,{=}\,\langle \hat{S}^{3}\rangle$ and
$\mu_8 \,{=}\, \langle \hat{Q}^{[2,0]}\rangle$ do not depend on the
spatial point~$\bi{x}_n$. These components are suitable to be
\emph{order parameters}. Evidently, $\mu_3$ describes a normalized
magnetization (a ratio of $z$-projection of magnetic moment to a
saturation magnetization), $\mu_8$ is similarly connected to a
quadrupole moment.

Now we briefly show that the proposed quantum model admits ordered
states. In the mean field approximation a partition function is
calculated by the formula
\[Z(\mu_3,\mu_8,T) = \Tr e^{-\frac{h_{\rm MF}}{kT}}, \] where $h_{\rm
MF}$ denotes the one-site Hamiltonian
\[
h_{\rm MF} = \textstyle -(J-\frac{1}{2}K) \mu_3 \hat{S}^3 -
\frac{1}{2}K \mu_8 \hat{Q}^{[2,0]}-\frac{4}{3}K.
\]
The mentioned mean field exists if self-consistent relations are
held, in other words, if the system
\begin{eqnarray*}
&\mu_3 = \langle \hat{S}^3 \rangle_{\rm MF} = \frac{\Tr \hat{S}^3
e^{-\frac{h_{\rm MF}}{kT}}}{\Tr e^{-\frac{h_{\rm MF}}{kT}}}, \\
&\mu_8 = \langle \hat{Q}^{[2,0]} \rangle_{\rm MF}= \frac{\Tr
\hat{Q}^{[2,0]} e^{-\frac{h_{\rm MF}}{kT}}}{\Tr e^{-\frac{h_{\rm
MF}}{kT}}}.
\end{eqnarray*}
has a solution. After calculation of the mean field averages one
obtains the self-consistent relations in the form
\begin{eqnarray*}
&\mu_3 =\frac{2\sinh \frac{(J-\frac{K}{2})
\mu_3}{kT}}{\exp\Bigl\{-\frac{\sqrt{3}\,K \mu_8}{2kT}\Bigr\}+ 2\cosh
\frac{(J-\frac{K}{2}) \mu_3}{kT}}, \\ &\mu_8 = \frac{2}{\sqrt{3}}
\frac{\cosh \frac{(J-\frac{K}{2})
\mu_3}{kT}-\exp\Bigl\{-\frac{\sqrt{3}\,K
\mu_8}{2kT}\Bigr\}}{\exp\Bigl\{-\frac{\sqrt{3}\,K
\mu_8}{2kT}\Bigr\}+ 2\sinh \frac{(J-\frac{K}{2}) \mu_3}{kT}}.
\end{eqnarray*}
Solutions of the system correspond to ordered states of the magnetic
system in question.

An evident solution is the paramagnetic state $(\mu_3\,{=}\,0,\,
\mu_8\,{=}\,0)$. All other solutions depend on a temperature $T$,
and the exchange integrals $J$ and $K$. Note, that we consider the
ferromagnetic interaction in preference: $J\,{>}\,0$. Nontrivial
solutions appear at temperatures low than the critical one $T_{\rm
crit} \,{=}\, \frac{2}{3k}(J\,{-}\,\frac{1}{2}K)$. As $K\,{<}\,0$
there exists a ferromagnetic state with the values $(\mu_3
\,{=}\,1,\,\mu_8\,{=}\,\frac{1}{\sqrt{3}})$ at zero temperature, and
a nematic state with the values $(\mu_3
\,{=}\,0,\,\mu_8\,{=}\,\frac{1}{\sqrt{3}})$ at zero temperature. As
$K\,{>}\,0$ there exist four nontrivial solutions: two ferromagnetic
states with the values $(\mu_3
\,{=}\,1,\,\mu_8\,{=}\,\frac{1}{\sqrt{3}})$ and $(\mu_3
\,{=}\,\frac{2}{3},\,\mu_8\,{=}\,\frac{-1}{2\sqrt{3}})$ at zero
temperature, and two nematic states with the values $(\mu_3
\,{=}\,0,\,\mu_8\,{=}\,\frac{-2}{\sqrt{3}})$ and $(\mu_3
\,{=}\,0,\,\mu_8\,{=}\,\frac{1}{\sqrt{3}})$ at zero temperature. The
same states are declared in \cite{Matveev,Nauciel}. The states
$(\mu_3 \,{=}\,1,\,\mu_8\,{=}\,\frac{1}{\sqrt{3}})$ and $(\mu_3
\,{=}\,0,\,\mu_8\,{=}\,\frac{-2}{\sqrt{3}})$ are stable. The problem
of transient processes in the mean field approach is discussed, for
example, in \cite{Nauciel}. The analysis of solutions of the
self-consistent relations proves that ordered states in the proposed
model exist.

In the sequel we deal with the case $J\,{=}\,K$, which corresponds
to the boundary between the ferromagnetic and the nematic regions
(see the phase diagram of the bilinear-biquadratic $s\,{=}\,1$ model
in \cite{Buchta}). In this case, the Hamiltonian \eref{BBH_SU2} and
its mean field approximation are SU(3)-invariant. The latter gets
the form
\begin{equation}\label{Hamilt3MF}
    \hat{\mathcal{H}}_{\rm MF}=-{\textstyle\frac{1}{2}}J z\sum_{n}\sum_{a}
    \hat{P}_{n}^{a}\langle
    \hat{P}_{n}^{a}\rangle -{\textstyle\frac{4}{3}}JNz =
    -{\textstyle\frac{1}{2}}J z\sum_{n}\sum_{a}
    \hat{P}_{n}^{a}\mu_a   -{\textstyle\frac{4}{3}}JNz.
\end{equation}

\subsection{Motion equations for large-scale fluctuations of mean field}
Return to the quantum $\mathrm{SU(3)}$-invariant spin model with the
Hamiltonian \eref{BBH_SU3}. The Heisenberg equation for an evolution
of $\hat{P}^a_n$ has the form
\begin{equation}\label{HeisenbergEq}
\rmi\hbar \frac{\rmd \hat{P}_n^a}{\rmd
t}=[\hat{P}_n^a,\hat{\mathcal{H}}].
\end{equation}
We suppose that the magnetic system is ordered, then we take an
average of equation~\eref{HeisenbergEq} over the Heisenberg (time
independent) coherent states
\begin{eqnarray*}
|\psi(n)\rangle = \frac{1}{\sqrt{N}} \Bigl( c_1 (n) |1\rangle +
c_{-1} (n)
|{-}1\rangle + c_0 (n) |0\rangle \Bigr), \\
|c_1|^2 + |c_{-1}|^2 + |c_0|^2 =1.
\end{eqnarray*}
Alternatively, one can take an average by means of the density
matrix. In the both cases we neglect correlations between
fluctuations of the quantum fields $\{\hat{P}_n^a\}_{a=1}^8$ at
distinct sites, that is
\begin{equation}\label{ApproxDegenOrb}
\langle \hat{P}^a_n \hat{P}^b_m \rangle \approx \langle \hat{P}^a_n
\rangle \langle \hat{P}^b_m \rangle = \mu_a(\bi{x}_n)
\mu_b(\bi{x}_m).
\end{equation}
An averaging of equation \eref{HeisenbergEq} results in the
following equation for $\mu_a(\bi{x}_n)$:
\begin{equation}\label{AvgEqMotion}
\hbar\frac{\partial \mu_a (\bi{x}_n)}{\partial t} = 2Jl^2 C_{abc}
\mu_b(\bi{x}_n) \Bigl(\mu_{c,xx}(\bi{x}_n) +
\mu_{c,yy}(\bi{x}_n)\Bigr),
\end{equation}
which is a Hamiltonian one with respect to the Lie-Poisson bracket.

In order to investigate large-scale fluctuations of the mean field
$\{\mu_a(\bi{x}_n)\}_{a=1}^8$, we consider a continuum space instead
of the discrete lattice. It can be achieved by the well-known
limiting process. In the case of $\mathrm{SU}(2)$-magnetic system
(only bilinear intereations are taken into account), this limiting
process underlies the macroscopic phenomenological theory of
magnetism~\cite{Herring}. The limiting process replaces quantum
operators by densities of their averages, which serve as dynamical
variables. In our case, we deal with the densities $M_a$ of averages
of the spin and quadrupole moments:
\begin{equation*}
M_a(x) = \sum_{n} \mu_a(\bi{x}_n)\, \delta(\bi{x},\bi{x}_n),\qquad
\delta(\bi{x},\bi{x}_n) = \left\{\begin{array}{ll} \frac{1}{V_0} &
\bi{x}_n \,{\in}\, U(\bi{x})
\\ 0 & \bi{x}_n \,{\not{\!\in}}\, U(\bi{x}),
\end{array}\right.
\end{equation*}
where $V_0$ denotes a physically infinitesimal region of the
lattice, and $U(\bi{x})$ is the infinitesimal neighborhood of
$\bi{x}$. The Lie-Poisson bracket for $\{M_a(\bi{x})\}$ is defined
by
\begin{equation*}
\{M_a(\bi{x}), M_b(\bi{y})\} = C_{abc}
M_c(\bi{x})\,\delta(\bi{x}-\bi{y}),
\end{equation*}
where $\delta(\bi{x})$ is the Dirac function. Since dimensionless
quantities are more suitable, we~introduce $\mu_a(\bi{x})\,{=}\, V_0
M_a(\bi{x})$ instead of $M_a(\bi{x})$. Then equation
\eref{AvgEqMotion} gets the form
\begin{eqnarray}
\hbar\frac{\partial \mu_a(\bi{x})}{\partial t} = \{\mathcal{H}_{\rm
eff},\, \mu_a(\bi{x})\} = V_0 C_{abc} \mu_b(\bi{x}) \frac{\delta
\mathcal{H}_{\rm eff}}{\delta \mu_c}, \label{muEqMotion} \\
\mathcal{H}_{\rm eff} =\frac{J}{l^{d-2}} \int \sum_a \Bigl\langle
\frac{\partial \mu_a}{\partial \bi{x}},\, \frac{\partial
\mu_a}{\partial \bi{x}} \Bigr\rangle \, \rmd^d \bi{x}, \nonumber
\end{eqnarray}
where $l$ is the lattice distance, and $d$ is the lattice dimension.
Note, that in the \mbox{2-dimensional} case we obtain a
scale-invariant Hamiltonian.

Evidently, \eref{muEqMotion} is a generalization of the well-known
Landau-Lifshitz equation to the case of 8-component vector field
$\{\mu_a\}$. In the same way one can obtain the standard
Landau-Lifshitz equation, if considers a spin system with
$s\,{=}\,\frac{1}{2}$ over the 2-dimensional space of representation
of SU(2).

We rewrite \eref{muEqMotion} in the matrix form
\begin{equation}\label{LLEq}
\hbar\frac{\partial \hat{\mu}}{\partial t} = \frac{2JV_0}{l^{d-2}}\,
[\hat{\mu},\Delta \hat{\mu}], \qquad \hat{\mu} = \sum_a \mu_a
\hat{P}^a.
\end{equation}
Here $\hat{\mu}$ is a Hermitian $3\times 3$ matrix, $[\cdot,\cdot]$
denotes the matrix commutator, $\Delta$ is the Laplas operator.
Being SU(3)-invariant equation  \eref{LLEq} as well as
\eref{muEqMotion} preserves the quantities $h_0 \,{=}\,
\frac{1}{2}\,\Tr \hat{\mu}^2$ and $f_0 \,{=}\, \frac{1}{2}\,\Tr
\hat{\mu}^3$, which we call invariants. They serve as constrains for
the Hamiltonian system and define the manifold where the vector
field $\{\mu_a\}$ lives. At the same time, this manifold is an orbit
of coadjoint representation of the group SU(3).

\section{Classical Hamiltonian systems on coadjoint orbits of SU(3)}

In the 1-dimensional case the Hamiltonian system \eref{muEqMotion}
appears to be integrable, what is shown below by means of the
orbital approach.

\subsection{Phase space for SU(3)-symmetric generalization
of Landau-Lifshitz equation}

In this section we briefly construct the orbital interpretation of a
finite-zone phase space for the $\mathrm{SU(3)}$-symmetric
generalization of the Landau-Lifshitz equation.

Consider an algebra of polynomials in $\lambda$ with coefficients
from the Lie algebra $\mathfrak{su}(3)$. Denote by
$\widetilde{\mathfrak{g}}_+$ the algebra $\mathfrak{su}(3)\otimes
\mathcal{P}(\lambda)$, where $\mathcal{P}(\lambda)$ is a ring of
polynomials in $\lambda$ with the standard multiplication. Let $A,\,
B \in \widetilde{\mathfrak{g}}_+$ have the form: $$ A(\lambda) =
\sum_{n=0}^{N+1} \hat{A}^n \lambda^n,\qquad B(\lambda) =
\sum_{k=0}^{N+1} \hat{B}^k \lambda^k,\quad \hat{A}^n,\, \hat{B}^k
\in \mathfrak{su}(3).$$ Then
\begin{equation}\label{GradOp}
[A,\,B] = \sum_{n,k} [\hat{A}^n,\,\hat{B}^k] \lambda^{n+k}\in
\widetilde{\mathfrak{g}}_+.
\end{equation}
The operation \eref{GradOp} turns $\widetilde{\mathfrak{g}}_+$ into
a graded Lie algebra.

Let $\hat{P}^{a,n} \,{=}\, \lambda^n \hat{P}^a$, where $a$ runs from
1 to 8. The set $\{\hat{P}^{a,n}\}$ serves as a basis in
$\widetilde{\mathfrak{g}}_+$. Recall that $[\hat{P}^a, \hat{P}^b] =
\rmi C_{abc} \hat{P}^c$; the nonzero components $C_{abc}$ have the
following values:
\begin{eqnarray*}
&C_{123}=C_{145}=C_{167}=C_{264}=C_{257}=C_{356}=1,\\
&C_{168}= C_{528}=\sqrt{3}, \quad C_{437}=2.
\end{eqnarray*}

Introduce a bilinear $\ad$-invariant form on
$\widetilde{\mathfrak{g}}_+$ by
\begin{equation}\label{BilinForm}
\langle A,B \rangle = \frac{1}{2}\res \lambda^{-N-2} \Tr A(\lambda)
B(\lambda).
\end{equation}
The basis $\{\hat{P}^{a,n}\}$ is orthonormal with respect to the
bilinear form. Let $\mathcal{M}=\widetilde{\mathfrak{g}}_{+}^{\ast}$
be a dual space to the algebra $\widetilde{\mathfrak{g}}_{+}$ with
respect to \eref{BilinForm}. Orthonormality of $\{\hat{P}^{a,n}\}$
implies that $\{\hat{P}^{a,n}\}$ also form a basis in $\mathcal{M}$.
Consider the following elements of $\mathcal{M}$:
\[
 \hat{\mu}(\lambda) = \sum_{n=0}^{N} \sum_{a=1}^8 \mu_a^n \lambda^n \hat{P}^a +
\bigl(\mu_3^{N+1} \hat{P}^3+ \mu_8^{N+1}
\hat{P}^8\bigr)\lambda^{N+1}.
\]
The functions $\hat{\mu}(\lambda)$ form a closed $\ad$-invariant
subset of $\mathcal{M}$, we denote it by $\mathcal{M}^{N+1}$. One
can compute the coordinate $\mu_a^n$ of $\hat{\mu}(\lambda)$ by the
formula
\[
\mu_a^n = \langle \hat{\mu}(\lambda), \hat{P}^{a,-n+N+1} \rangle.
\]

Define a Lie-Poisson bracket in $\mathcal{C}(\mathcal{M}^{N+1})$ as
\begin{equation}\label{Lie-PoissonBracket}
  \{f_1,f_2\} = \sum_{m,n} \sum_{a,b}^{8}
  W_{ab}^{mn} \frac{\partial f_1}{\partial \mu_a^m}
  \frac{\partial f_2}{\partial \mu_b^n}
\end{equation}
with the Poisson tensor field
\[W_{ab}^{mn}=\langle \hat{\mu}(\lambda),
  [\hat{P}^{a,-m+N+1},\hat{P}^{b,-n+N+1}]\rangle.\]
  Introduce also two $\ad$-invariant functions $I_2(\lambda)$ and
$I_3(\lambda)$ by the formulas
\begin{eqnarray*}
I_2(\lambda) = \frac{1}{2}\Tr \hat{\mu}^2(\lambda)= \sum_{a} \mu_a^2(\lambda),\\
I_3(\lambda) = \frac{1}{2}\Tr \hat{\mu}^3(\lambda) =
\textstyle\sqrt{\frac{5}{3}}\,d_{abc}\mu_a(\lambda)\mu_b(\lambda)\mu_c(\lambda),
\end{eqnarray*}
where $d_{abc}=\frac{\sqrt{3}}{4\sqrt{5}}\Tr (\hat{P}^a \hat{P}^b
\hat{P}^c + \hat{P}^b \hat{P}^a \hat{P}^c)$, and $\mu_a (\lambda)$
denotes the polynomial
\[\mu_a (\lambda) = \mu_a^0 + \mu_a^1 \lambda + \mu_a^2 \lambda^2 +
\cdots + \mu_a^{N+1} \lambda^{N+1}.\]

The invariant functions are also polynomials in $\lambda$:
\begin{eqnarray*}
I_2(\lambda) = h_0 + h_1\lambda + \cdots + h_{2N+2} \lambda^{2N+2},
\\ I_3(\lambda) = f_0 + f_1\lambda + \cdots + f_{3N+3}
\lambda^{3N+3}.
\end{eqnarray*}
It is easy to prove that the coefficients $\{h_{0}$, $\dots$,
$h_{N+1}$, $f_{0}$, $\dots$, $f_{N+1}\}$ are annihilators with
respect to the bracket \eref{Lie-PoissonBracket}. We fix these
coefficients and obtain the system of algebraic equations
\begin{equation}\label{OrbitEq}
  h_{n}={\rm const},\ f_{n}={\rm const},\
  n=0,\,\dots,\, N+1,
\end{equation}
which determines an embedding of an orbit $\mathcal{O}^{N+1}$ of
dimension $6(N{+}1)$ into $\mathcal{M}^{N+1}$. The coefficients
$\{h_{N+2}$, $\dots,$ $h_{2N+2}$, $f_{N+2}$, $\dots,$ $f_{3N+3}\}$
are pairwise commutative integrals of motion. We call them
Hamiltonians. In the 1-dimensional case the number of Hamiltonians
is sufficient for integrability of the Hamiltonian system on an
orbit.

Here we are interested in two Hamiltonians: $h_{N+2}$, $h_{N+3}$,
and the corresponding Hamiltonian flows. The Hamiltonian $h_{N+2}$
gives rise to the stationary flow
\begin{equation}\label{StatFlow}
\frac{\partial \mu_a^{n}}{\partial x} = \{\mu_a^{n}, h_{N+2}\} =
2C_{abc}\mu_b^0 \mu_c^{n+1},\quad a=1,\,\dots,\,8.
\end{equation}
The Hamiltonian $h_{N+3}$ gives rise to the evolutionary flow
\begin{equation}\label{EvolFlow}
\frac{\partial \mu_a^n}{\partial t} = \{\mu_a^{n}, h_{N+3}\} =
2C_{abc}\bigl(\mu_b^0 \mu_c^{n+2}+\mu_b^1 \mu_c^{n+1}\bigr),\quad
a=1,\,\dots,\,8.
\end{equation}

Equations \eref{StatFlow} and \eref{EvolFlow} are compatible, for
the corresponding Hamiltonians commute: $\{h_{N+2},\,
h_{N+3}\}\,{=}\,0$. Thus, \eref{EvolFlow} describes an evolution on
the trajectories of \eref{StatFlow}, that is the dynamical variables
$\{\mu_a^n\}$ in \eref{EvolFlow} depend on $x$. From \eref{StatFlow}
and \eref{EvolFlow} we have:
\begin{equation}\label{Nullcurvature}
\frac{\partial \mu_a^0}{\partial t} = 2C_{abc}\mu_b^0 \mu_c^{2}=
\frac{\partial \mu_a^1}{\partial x}.
\end{equation}
The variables $\{\mu_a^1\}$ can be expressed in the terms of
$\{\mu_a^0\}$ and $\{\frac{\partial}{\partial x}\mu_a^0\}$, then
\eref{Nullcurvature} becomes a closed system of partial equations
for $\{\mu_a^0\}$. In order to compute the variables $\{\mu_a^1\}$
one has to solve the following degenerate system of equations of the
stationary flow:
\begin{equation}\label{StatFlowDegen}
\frac{\partial \mu_a^0}{\partial x} = 2C_{abc}\mu_b^0
\mu_c^{1},\quad a=1,\,\dots,\,8.
\end{equation}
It becomes possible if one restricts the system to the orbit
$\mathcal{O}^{N+1}\,{\subset}\, \mathcal{M}^{N+1}$.

\subsection{Classification of orbits of SU(3)}
It is evident, that the orbit $\mathcal{O}^{N+1}$ defined by
\eref{OrbitEq} is a vector bundle over a coadjoint orbit of the
group SU(3). That is why we need to classify orbits of SU(3).

The group SU(3) is simple~\cite{Helgason}, hence its algebra
$\mathfrak{g}\,{\simeq}\,\mathfrak{su}(3)$ coincides with the dual
space~$\mathfrak{g}^{\ast}$. Consequently, the coordinates
$\{\mu_a\}$ in $\mathfrak{g}^{\ast}$ can be regarded as coordinates
in $\mathfrak{su}(3)$ as well as in $\mathfrak{su}^{\ast}(3)$. A
generic element $\hat{\mu}\,{\in}\, \mathfrak{su}^{\ast}(3)$ has the
 form
\begin{equation}\label{Melement} \small
\fl \hat{\mu}= \left(\begin{array}{ccc}
\mu_3+\frac{1}{\sqrt{3}}\mu_8 & \mu_7-\rmi\mu_4& \frac{1}{\sqrt{2}}(\mu_1-\rmi\mu_6+\mu_5-\rmi\mu_2) \\
\mu_7+\rmi\mu_4 & -\mu_3+\frac{1}{\sqrt{3}}\mu_8& \frac{1}{\sqrt{2}}(\mu_1-\rmi\mu_6-\mu_5+\rmi\mu_2) \\
\frac{1}{\sqrt{2}}(\mu_1+\rmi\mu_6+\mu_5+\rmi\mu_2)&\frac{1}{\sqrt{2}}(\mu_1+\rmi\mu_6-\mu_5-\rmi\mu_2)&
-\frac{2}{\sqrt{3}}\mu_8
\end{array}\right)\!\!.
\end{equation}
Let $\mathfrak{h}$ be the maximal commutative subalgebra (also
called the Cartan subalgebra) of $\mathfrak{g}$. The dual space
$\mathfrak{h}^{\ast}$ to the Cartan subalgebra $\mathfrak{h}$
coincides with  $\mathfrak{h}$.

By definition the set $\mathcal{O}_{\hat{\mu}_{\rm in}} \,{=}\,
\{g^{-1} \hat{\mu}_{\rm in} g,\, \forall g\,{\in}\,
\mathrm{SU}(3)\}$ is the \emph{coadjoint orbit} of SU(3) through an
\emph{initial point} $\hat{\mu}_{\rm in}\,{\in}\,
\mathfrak{su}^{\ast}(3)$. All elements $g'\,{\in}\, \mathrm{SU}(3)$
such that $g'{}^{-1} \hat{\mu}_{\rm in} g' \,{=}\, \hat{\mu}_0$ form
the stationary subgroup $\mathrm{S}_{\hat{\mu}_{\rm in}}$ at
$\hat{\mu}_{\rm in}$. The orbit $\mathcal{O}_{\hat{\mu}_{\rm in}}$
is a homogeneous space, which is diffeomorphic to the coset space
$\mathrm{SU}(3) / \mathrm{S}_{\hat{\mu}_{\rm in}}$. There exist two
types of orbits of SU(3): the \emph{generic} $\mathcal{O}_{\rm
gen}\,{=}\,\frac{\mathrm{SU}(3)}{\mathrm{U}(1)\times \mathrm{U}(1)}$
of dimension 6, and the \emph{degenerate} $\mathcal{O}_{\rm
deg}\,{=}\,\frac{\mathrm{SU}(3)}{\mathrm{SU}(2)\times
\mathrm{U}(1)}$ of dimension~4.

It is proven by R.~Bott that \emph{each orbit of coadjoint action of
a semisimple group G intersects $\mathfrak{h}^{\ast}$ precisely in
an orbit of the Weyl group W(G)}.

The full Weyl group of SU(3) consists of six elements $\{e,\,
\sigma_1,\, \sigma_2,\, \sigma_1\sigma_2,\, \sigma_2\sigma_1,\,
\sigma_1\sigma_2\sigma_1\}$, where $\sigma_1$, $\sigma_2$ are
reflections across the hyperplanes orthogonal to the simple roots
$\alpha_1$, $\alpha_2$ (see figure~1). The open domain $C \,{=}\,
\{\hat{\mu}\,{\in}\, \mathfrak{h}^{\ast},\, \langle \hat{\mu},
\alpha \rangle \,{>}\, 0, \forall \alpha\,{\in}\, \Delta^{+}\}$ is
called a positive Weyl chamber. Here $\Delta^{+}$ denotes the set of
positive roots. We call the set
$\Gamma_{\alpha}\,{=}\,\{\hat{\mu}\,{\in}\, \mathfrak{h}^{\ast},\,
\langle \hat{\mu}, \alpha \rangle \,{=}\, 0\}$ a wall of the Weyl
chamber. An orbit of the Weyl group $\mathrm{W(G)}$ is obtained by
the action of $\mathrm{W(G)}$ on a point of $\overline{C}$.
\begin{figure}[h]
\centering
\includegraphics[width=0.37\textwidth]{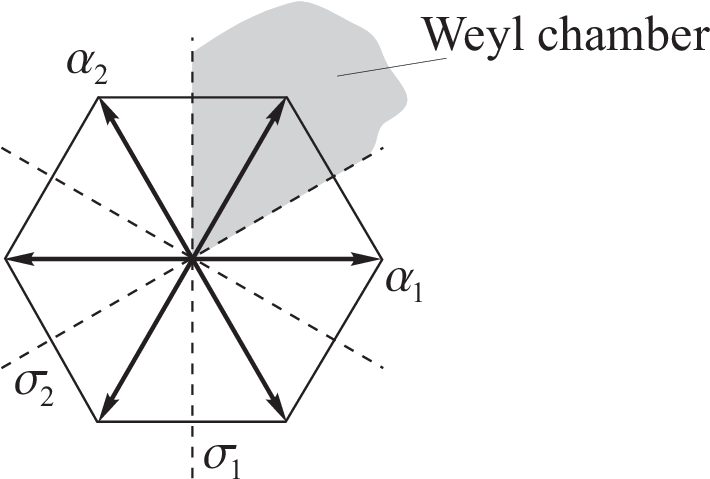}
\caption{Root diagram of SU(3).}
\end{figure}

\emph{Each orbit of the Weyl group W(G), and consequently, each
coadjoint orbit of  G intersects the positive Weyl chamber in an
only point.} That is why we can classify coadjoint orbits of
$\mathrm{G}$ by points of the positive Weyl chamber.

In the case of group SU(3), there exist two types of orbits of the
Weyl group. A~generic orbit contains six elements and passes through
the interior of the positive Weyl chamber. A degenerate orbit
contains three elements and passes through a wall of the positive
Weyl chamber. According to this, we call an orbit of SU(3) a
\emph{generic} one if $\hat{\mu}_{\rm in}$ lies in the interior of
the positive Weyl chamber, and a \emph{degenerate} one if
$\hat{\mu}_{\rm in}$ belongs to a wall of the positive Weyl chamber.

In our case, $\hat{\mu}_{\rm in}$ has the following diagonal form
\[\hat{\mu}_{\rm in} = \diag(m+{\textstyle\frac{1}{\sqrt{3}}}\,q,\,
-m+{\textstyle\frac{1}{\sqrt{3}}}\,q,\,
-{\textstyle\frac{2}{\sqrt{3}}}\,q ),\] where $m$ and $q$ denote
initial values of the variables $\mu_3$ and $\mu_8$ respectively, or
boundary values (at zero temperature) of the corresponding
components of the mean field. As~$m\,{>}\,0$, $q\,{>}\,0$ the
coadjoint action of SU(3) gives a generic orbit. If $m\,{=}\,0$ or
$m\,{=}\,\sqrt{3}\,q$, we obtain a degenerate orbit. In the sequel
we consider degenerate orbits with $m\,{=}\,0$.

\subsection{Hamiltonian equations on orbits of SU(3)}
Return to the system of equations \eref{StatFlowDegen}, which is
degenerate in  $\mathcal{M}^{N+1}$. However, it can be solved if one
restricts the system to the orbit
$\mathcal{O}^{N+1}\,{\subset}\,\mathcal{M}^{N+1}$. Each orbit is
determined by the following equation~\cite{Skrypnik}:
\begin{equation}\label{Constr}
\chi_{\rm min}(\hat{\mu}) = 0,
\end{equation}
where $\chi_{\rm min}(\hat{\mu})$ is the minimal characteristic
polynomial in $\hat{\mu}\,{\in}\, \mathcal{O}^{N+1}$. Equation
\eref{Constr} serves as a constrain for the system
\eref{StatFlowDegen}, which has the form
\begin{equation}\label{AdEq}
\frac{\partial \hat{\mu}^0}{\partial x} = \Ad_{\hat{\mu}^0}
\hat{\mu}^1.
\end{equation}
Now we solve \eref{AdEq} on orbits of the group SU(3).

A degenerate orbit is determined by the equation
\begin{equation*}
\hat{\mu}^2 + \textstyle\sqrt{\frac{h_0}{3}}\,\hat{\mu} -
\frac{2h_0}{3} = 0,
\end{equation*}
where $h_0 \,{=}\, q^2 \,{=}\, \textrm{const}$. Using this
constrain, one obtains the following solution of~\eref{AdEq}:
$\mu^1_a \,{=}\, \frac{1}{6h_0}\, C_{abc} \mu_b^0 \mu_{c,x}^0
\,{+}\, \frac{h_1}{2h_0}\,\mu_{a}^0$, where
$\frac{h_1}{2h_0}\,\mu_{a}^0$ is an element of $\Ker
\Ad_{\hat{\mu}^0}$. The motion equation \eref{Nullcurvature} on the
degenerate orbit has the form
\begin{equation}\label{MotionEqsDeg}
\frac{\partial \mu_a}{\partial t} =
\textstyle\frac{8\mathcal{A}}{3h_0}\, C_{abc} \mu_b \mu_{c,xx} +
\frac{8\mathcal{A} h_1}{h_0}\,\mu_{a,x},
\end{equation}
where we write $\mu_a$ instead of $\mu^0_a$ and scale the flow
parameter $t$ by $16\mathcal{A}$. The dimensional constant
$\mathcal{A}$ provides a correspondence between \eref{MotionEqsDeg}
as $h_1\,{=}\,0$ and  \eref{LLEq} as $d\,{=}\,1$. That is
\eref{MotionEqsDeg} describes large-scale fluctuations of the mean
field $\{\mu_a\}$.

A generic orbit  is determined by the characteristic equation
\begin{equation*}
\hat{\mu}^3 - h_0 \hat{\mu} - \textstyle\frac{2}{3}f_0 = 0,
\end{equation*}
where $h_0 \,{=}\, m^2\,{+}\,q^2$, and $f_0 \,{=}\,
\frac{1}{\sqrt{3}}\bigl(3m^2q\,{-}\,q^3\bigr)$. On this orbit we
obtain the following solution of \eref{AdEq}:
\begin{eqnarray*}
\mu_a^1 &= \textstyle\frac{1}{8(h_0^3-3 f_0^2)} \left(h_0^2 C_{abc}
\mu^0_b \mu^0_{c,x}  - 2\sqrt{3}\,f_0 C_{abc} \eta^0_b \mu^0_{c,x} +
h_0 C_{abc} \eta^0_b \eta^0_{c,x} \right) + \\ &+
\textstyle\frac{2f_0f_1-3 h_0^2
 h_1}{6(f_0^2-h_0^3)}\mu^0_{a}
+ \frac{3 f_0 h_1-2 h_0 f_1}{6\sqrt{3}\,(f_0^2-h_0^3)}\eta^0_{a},
\end{eqnarray*}
where $\eta^0_a = \sqrt{5}\, d_{abc} \mu_b^0 \mu_{c}^0$. The motion
equation \eref{Nullcurvature} on the generic orbit has the form
\begin{eqnarray}\label{MotionEqsGen}
\frac{\partial \mu_a}{\partial t} &=
\textstyle\frac{2\mathcal{A}}{h_0^3-3 f_0^2}\Bigl(h_0^2 C_{abc}\mu_b
\mu_{c,xx} - \sqrt{3}\,f_0 C_{abc} \mu_b \eta_{c,xx}  - \nonumber\\
&-\sqrt{3}\, f_0 C_{abc} \eta_{b}
 \mu_{c,xx} + h_0 C_{abc} \eta_b \eta_{c,xx}  \Bigr)+ \\ &+ \textstyle
 \frac{8\mathcal{A}}{3}\frac{2f_0f_1-3 h_0^2
 h_1}{f_0^2-h_0^3}\,\mu_{a,x}
+\frac{8\mathcal{A}}{3\sqrt{3}} \frac{3f_0 h_1-2 h_0
f_1}{f_0^2-h_0^3}\,\eta_{a,x}. \nonumber
\end{eqnarray}
As $h_1\,{=}\,0$, $f_1\,{=}\,0$ equation \eref{MotionEqsGen} also
describes large-scale fluctuations of the mean field. One can obtain
\eref{MotionEqsGen} from \eref{HeisenbergEq} by averaging with a
more complicate correlation rule.

Equations \eref{MotionEqsDeg} and \eref{MotionEqsGen} imply the
following Hamiltonians respectively:
\begin{eqnarray*}
&\mathcal{H}_{\rm deg} = \textstyle \frac{4\mathcal{A}/\hbar}{3h_0}
\displaystyle \int \sum_{a=1}^8 (\mu_{a,x})^2\, \rmd x,\\
&\mathcal{H}_{\rm gen} = \textstyle \frac{\mathcal{A}/\hbar}{h_0^3-3
f_0^2} \displaystyle \int \sum_{a=1}^8 \Bigl( h_0^2 (\mu_{a,x})^2 +
h_0 (\eta_{a,x})^2 - 2\sqrt{3}\,f_0 \mu_{a,x}\eta_{a,x} \Bigr)\,
\rmd x.
\end{eqnarray*}

In addition to the 1-dimensional  case one can consider the
corresponding 2- or 3-dimensional Hamiltonian systems with the
effective Hamiltonians
\begin{eqnarray}\label{HamEff}
\mathcal{H}^{\rm eff} = \mathcal{J} \int H(\bm{\mu})\, \rmd^d
\bi{x},
\end{eqnarray}
where $\bm{\mu}$ denotes $\{\mu_a\}_{a=1}^8$. The exchange integral
$\mathcal{J}\,{=}\,\mathcal{A}/\hbar$ gives the Hamiltonian the
required physical dimension. By $H$ we denote the Hamiltonian
density
\begin{eqnarray*}
H_{\rm deg} = \textstyle \frac{4}{3h_0} \displaystyle \sum_{k=1}^d
 \sum_{a=1}^8 (\mu_{a,x_k})^2,\quad \textup{or} \\ H_{\rm gen} = \textstyle
\frac{1}{h_0^3-3 f_0^2} \displaystyle \sum_{k=1}^d \sum_{a=1}^8
\Bigl( h_0^2 (\mu_{a,x_k})^2 + h_0 (\eta_{a,x_k})^2 - 2\sqrt{3}\,f_0
\mu_{a,x_k}\eta_{a,x_k} \Bigr).
\end{eqnarray*}
One can use these effective Hamiltonians for describing the magnetic
system considered in section 2. Note, that $\mathcal{H}_{\rm deg}$
is the same as the Hamiltonian of \eref{muEqMotion}.

The proposed Hamiltonians describe large-scale (slow) fluctuations
of the mean field~$\bm{\mu}$. After averaging over high frequencies
some observed quantities become rigid (or invariant); these
quantities are $h_0\,{=}\,\delta_{ab}\mu_a \mu_b$, and
$f_0\,{=}\,\sqrt{5/3}\, d_{abc}\mu_a\mu_b\mu_c$. They serve as
constrains for the Hamiltonian systems, and are equivalent to
\eref{Constr}. The constrains determine the orbit where the system
has to be considered.

In the case of SU(3)-invariant model we deal with the magnet whose
ferromagnetic and nematic states are equiprobable. A generic orbit
corresponds to a state with the ferromagnetic order at zero
temperature, because of nonzero magnetization ($m\,{\neq}\, 0$).
A~degenerate orbit ($m\,{=}\,0$) corresponds to a state with the
nematic order at zero temperature. So equations \eref{MotionEqsDeg}
and \eref{MotionEqsGen} describe  fluctuations of the mean field
$\bm{\mu}$ near a nematic and a ferromagnetic ordered states
respectively.

\subsection{SU(3)-invariance of effective Hamiltonians}
As mentioned in Section 2, the quantum Hamiltonian \eref{BBH_SU2}
and the mean field Hamiltonian \eref{HamiltMidField} are
SU(3)-invariant as $K\,{=}\,J$. Here we show that the proposed
classical effective Hamiltonians \eref{HamEff} are also
SU(3)-invariant.

Recall, that the mean field $\{\mu_a\}$ belongs to the real
8-dimensional space of coadjoint representation of SU(3). Hence, an
action of SU(3) transforms $\{\mu_a\}$ by the formula
\begin{equation*}
  \mu_{a} = \hat{D}_{ab}\mu_{b},\quad  \hat{D}_{ab} \in
  \mathrm{SO}(8),
\end{equation*}
$\hat{D}_{ab}$ is a matrix of the real irreducible 8-dimensional
representation of the group SU(3).

Note, that the tensor $d_{abc}$ satisfies the relation $d_{abc}
d_{qbc}\,{=}\,\delta_{aq}$. The components $\{d_{abc}\}$ serve as
Clebsch-Gordon coefficients for a decomposition of tensor square of
the coadjoint representation into irreducible components.  In this
connection, we have the following relation, well-known in theory of
representations, $\hat{D}_{bb'}\hat{D}_{cc'} \,{=}\,
d_{qbc}d_{q'b'c'} \hat{D}_{qq'}$. Then as a result of the action of
 SU(3) on $\{\eta_a\}$ we get
\begin{equation*}
  \eta_a = \sqrt{5}\, d_{abc} \hat{D}_{bb'}\mu_{b'} \hat{D}_{cc'}\mu_{c'} =
  \hat{D}_{qq'} \eta_{q'}.
\end{equation*}
The action of SU(3) on the vector fields $\{\mu_{a,x}\}$ and
$\{\eta_{a,x}\}$ is the same. Therefore, the densities $H_{\rm gen}$
and $H_{\rm deg}$ are SU(3)-invariant.

Densities of the effective Hamiltonians can be expressed as
\begin{equation}\label{SigmaDens}
  H = \sum_{jk} \sum_{ab} g_{ab}(\bm{\mu}) \frac{\partial \mu_a}{\partial x_j}
  \frac{\partial \mu_b}{\partial x_k}\,  G_{jk}(\bi{x}),
\end{equation}
where $g_{ab}(\bm{\mu})$ serves as a metrics invariant under an
action of the group that transforms~$\bm{\mu}$, and $G_{jk}(\bi{x})$
is a metrics in the $\bi{x}$-space. For the proposed effective
Hamiltonians the $\bi{x}$-space is Euclidean: $G_{jk}(\bi{x}) =
\delta_{jk}$. The metrics in $\bm{\mu}$-space is trivial:
$g_{ab}(\bm{\mu}) \,{=}\, \frac{4}{3h_0}\,\delta_{ab}$ in the case
of a degenerate orbit, and has a more complicate form:
$$g_{ab}(\bm{\mu}) = \textstyle\frac{1}{h_0^3-3 f_0^2} \Bigl( h_0^2
\delta_{ab} + 20 h_0 d_{cpa} d_{cqb} \mu_p \mu_q  - 4\sqrt{15}\,f_0
d_{abc}\mu_c \Bigr)$$ in the case of a generic orbit.

The density \eref{SigmaDens} can be interpreted as a Lagrangian
density of relativistic $\sigma$-model; in this case $G_{jk}$ is the
metrics of the Minkowski space. After quantization one obtains a
Hamiltonian system that describes slow fluctuations. Quick
fluctuations can be taken into account  by means of a
renormalization group~\cite{Tsvelyk}. It makes the coefficients
$\frac{1}{h_0^3-3f_0^2} $ and $\frac{4}{3h_0}$ dependent on
parameters of the renormalization group, for example on a
temperature.

\subsection{Parametrization of orbits}
Remarkably, that the effective models are entirely defined by
geometry of orbits. We will prove this statement, if perform a
parametrization of orbits and express the effective Hamiltonians in
terms of these parameters.

A generalized stereographic projection gives a suitable way of
parametrization for coadjoint orbits of a semisimple Lie group
\cite{Bernatska}. In the case of group SU(3) we have
\begin{equation*}
\mu_a = -\textstyle\frac{m-\sqrt{3}\, q}{2}\, \zeta_a + m \xi_a,
\qquad \eta_a = \frac{\sqrt{3}(m^2-q^2)-2mq}{2}\, \zeta_a + 2mq
\xi_a,
\end{equation*}
where
\begin{equation*}\label{ParamVar}
\begin{array}{ll}
\zeta_1 = -\frac{1}{\sqrt{2}}
\frac{w_2+w_3+\bar{w}_2+\bar{w}_3}{1+|w_2|^2+|w_3|^2} &\xi_1 =
-\frac{1}{\sqrt{2}}\frac{(1-w_1)(\bar{w}_3-\bar{w}_1\bar{w}_2)
+(1-\bar{w}_1)(w_3-w_1w_2)}{1+|w_1|^2+|w_3-w_1w_2|^2},\\
\zeta_2= \frac{-i}{\sqrt{2}}
\frac{w_2-w_3-\bar{w}_2+\bar{w}_3}{1+|w_2|^2+|w_3|^2} & \xi_2 =
\frac{-i}{\sqrt{2}}\frac{(1+w_1)(\bar{w}_3-\bar{w}_1\bar{w}_2)
-(1+\bar{w}_1)(w_3-w_1w_2)}{1+|w_1|^2+|w_3-w_1w_2|^2},\\
\zeta_3 = \frac{|w_2|^2-|w_3|^2}{1+|w_2|^2+|w_3|^2}
& \xi_3 = \frac{1-|w_1|^2}{1+|w_1|^2+|w_3-w_1w_2|^2},\\
\zeta_4 = i\frac{\bar{w}_2w_3-w_2\bar{w}_3}{1+|w_2|^2+|w_3|^2} &
\xi_4 = i
\frac{w_1-\bar{w}_1}{1+|w_1|^2+|w_3-w_1w_2|^2},\\
\zeta_5 = \frac{1}{\sqrt{2}}
\frac{w_2-w_3+\bar{w}_2-\bar{w}_3}{1+|w_2|^2+|w_3|^2} & \xi_5 = -
\frac{1}{\sqrt{2}}\frac{(1+w_1)(\bar{w}_3-\bar{w}_1\bar{w}_2)
+(1+\bar{w}_1)(w_3-w_1w_2)}{1+|w_1|^2+|w_3-w_1w_2|^2},  \\
\zeta_6 = \frac{i}{\sqrt{2}}
\frac{w_2+w_3-\bar{w}_2-\bar{w}_3}{1+|w_2|^2+|w_3|^2} & \xi_6 =
\frac{i}{\sqrt{2}}\frac{(1-\bar{w}_1)(w_3-w_1w_2)-(1-w_1)(\bar{w}_3-\bar{w}_1\bar{w}_2)}
{1+|w_1|^2+|w_3-w_1w_2|^2},\\
\zeta_7 = -\frac{\bar{w}_2w_3+w_2\bar{w}_3}{1+|w_2|^2+|w_3|^2} &
\xi_7 = -\frac{w_1+\bar{w}_1}{1+|w_1|^2+|w_3-w_1w_2|^2},\\
\zeta_8
=\frac{1}{\sqrt{3}}\frac{2-|w_2|^2-|w_3|^2}{1+|w_2|^2+|w_3|^2} &
\xi_8 = \frac{1}{\sqrt{3}}
\frac{1+|w_1|^2-2|w_3-w_1w_2|^2}{1+|w_1|^2+|w_3-w_1w_2|^2}.
\end{array}
\end{equation*}
Here $w_1$, $w_2$, $w_3$ are complex parameters on a generic orbit,
$m$ and $q$ are initial values of $\mu_3$ and $\mu_8$ respectively.
The initial values fix an orbit.  For a degenerate orbit one has to
assign $m=0$ and $w_1=0$.

After this parameterization the effective Hamiltonians get the form
\begin{eqnarray*}
\mathcal{H}^{\rm eff} = \int \sum_{k=1}^d \sum_{\alpha,\,\beta}
g_{\alpha\beta}(\bi{w})\, \frac{\partial w_{\alpha}}{\partial x_k}
\frac{\partial w_{\beta}}{\partial x_k}\, \rmd^d \bi{x}, \\ g^{\rm
deg}_{\alpha\beta}= \sum_{a} \frac{\partial \zeta_a}{\partial
w_{\alpha}}\frac{\partial \zeta_a}{\partial w_{\beta}},\\ g^{\rm
gen}_{\alpha\beta}= \sum_{a} \left( \frac{\partial \zeta_a}{\partial
w_{\alpha}}\frac{\partial \zeta_a}{\partial w_{\beta}} -
\frac{\partial \zeta_a}{\partial w_\alpha}\frac{\partial
\xi_a}{\partial w_\beta} + \frac{\partial \xi_a}{\partial
w_\alpha}\frac{\partial \xi_a}{\partial w_\beta} \right).
\end{eqnarray*}
The tensors $g^{\rm gen}$ and $g^{\rm deg}$ serve as metrics on
orbits in terms of the complex parameters $\bi{w} \,{=}\,
\{w_1,\,\bar{w}_1,\,w_2,\,\bar{w}_2,\,w_3,\, \bar{w}_3\}$ for a
generic orbit, and $\bi{w} \,{=}\, \{w_2,\,\bar{w}_2,\,w_3,\,
\bar{w}_3\}$ for a degenerate orbit. Note, that the metrics do not
depend on the initial values $m$ and~$q$, fixing an orbit. All
generic orbits have the same metrics, as well as degenerate orbits.

\section{Results and discussion}

Our main result is the following. For a magnetic system with the
spin $s\,{\geqslant}\,1$ we propose two effective classical models
that describe fluctuations of the mean field by the Landau-Lifshitz
like equations. We consider the 8-component mean field
$\bm{\mu}\,{=}\,\{\mu_a\}_{a=1}^8$, taking into account not only
magnetization but also quadrupole moments.

The effective models deal with large-scale (slow) fluctuations of
the mean field. Small-scale (quick) fluctuations are cut off by
quasiaveraging. In this process some observed quantities become
rigid and serve as constrains determining the manifold where the
mean field lives. This manifold appears to be a coadjoint orbit of
the group SU(3).

Also we propose a complex parametrization for the manifold and
reduce the mean field and the Hamiltonian density to complex
parameters. Remarkably, that in terms of the complex parameters the
density becames independent on boundary values of $\bm{\mu}$.
Moreover, the Hamiltonian density serve as a Riemannian metrics on
the manifold.

In the case of SU(3)-invariant model we deal with the magnet whose
ferromagnetic and nematic states at zero temperature are
equiprobable. That is why we propose two effective Hamiltonians:
$\mathcal{H}_{\rm gen}$ for states with the ferromagnetic order at
zero temperature, and $\mathcal{H}_{\rm deg}$ for states with the
nematic order (when magnetization is zero) at zero temperature. Also
we produce equations \eref{MotionEqsDeg} and \eref{MotionEqsGen}
describing large-scale fluctuations of the mean field $\bm{\mu}$
near a nematic and a ferromagnetic ordered states respectively.

The proposed classical models can be used to construct topological
excitations \cite{BernHolod}, which are stationary solutions of the
Landau-Lifshitz like equations. These excitations realize
destruction of a long-range order in 2-dimensional spin systems at
nonzero temperatures, according to the Mermin-Wagner theorem.

The considered scheme is easily extended to the case with higher
powers of exchange interaction. For an arbitrary spin~$s$ the spin
operators $\{\hat{S}_n^{\alpha}\}$ are defined over the
$(2s{+}1)$-dimensional space of representation of the group
$\mathrm{SU}(2)$. The complete matrix algebra generated by the spin
operators is $\Mat_{(2s{+}1)\times (2s{+}1)}$. Then one can consider
a spin Hamiltonian with powers of exchange interaction up to $2s$.
Such Hamiltonian admits a bilinear form, if one takes into account
multipole moments. In the mean field approximation this quantum
model corresponds to a Hamiltonian system on a coadjoint orbit of
the group $\mathrm{SU}(2s{+}1)$. Each orbit has a Hamiltonian
system, which serves as an effective classical model.

\section*{Acknowledgements}

The research presented in the paper was conducted with the financial
support of the Julian Wynnyckyj professorship in natural sciences at
Kyiv-Mohyla Academy, and the grant of the International Charitable
Fund for Renaissance of Kyiv-Mohyla Academy.

\end{document}